\documentclass[prd,aps,tightline,twocolumn,nofootinbib,superscriptaddress]{revtex4}

\usepackage{amssymb}
\usepackage{amsmath}
\usepackage[dvips]{graphicx}
\usepackage{latexsym}
\usepackage{bm}

\begin{document}

\begin{flushleft}
STUPP-11-208, MISC-2011-04, KEK-TH-1455, KEK-Cosmo-71,
 MP-20
\end{flushleft}
\hfill \today

\title{ 
Big-bang nucleosynthesis with a long-lived charged massive particle
including $^4$He spallation processes}

\author{Toshifumi Jittoh}
\affiliation{Department of Physics, Saitama University, 
     Shimo-okubo, Sakura-ku, Saitama, 338-8570, Japan}
     
\author{Kazunori Kohri}
\affiliation{Theory Center, Institute of Particle and Nuclear Studies,
KEK (High Energy Accelerator Research Organization),
1-1 Oho, Tsukuba 305-0801, Japan}
\affiliation{Department of Physics, Tohoku University, Sendai 980-8578, Japan}

\author{Masafumi Koike} 
\affiliation{Department of Physics, Saitama University, 
     Shimo-okubo, Sakura-ku, Saitama, 338-8570, Japan}

\author{Joe Sato}
\affiliation{Department of Physics, Saitama University, 
     Shimo-okubo, Sakura-ku, Saitama, 338-8570, Japan}
     
\author{Kenichi Sugai}
\affiliation{Department of Physics, Saitama University, 
     Shimo-okubo, Sakura-ku, Saitama, 338-8570, Japan}     

\author{Masato Yamanaka}
\affiliation{Theory Center, Institute of Particle and Nuclear Studies,
KEK (High Energy Accelerator Research Organization),
1-1 Oho, Tsukuba 305-0801, Japan}
\affiliation{Maskawa Institute for Science and Culture, Kyoto Sangyo University, 
Kyoto 603-8555, Japan}

\author{Koichi  Yazaki}
\affiliation{Hashimoto Mathematical Physics Laboratory , Nishina Accelerator Research Center, 
RIKEN, Wako, Saitama 351-0198, Japan}
\affiliation{Yukawa Institute for Theoretical Physics,Kyoto University, Kyoto 606-8502, Japan}


\begin{abstract}
We propose helium-4 spallation processes induced by long-lived 
stau in supersymmetric standard models, and investigate an 
impact of the processes on light elements abundances.  
We show that, as long as the phase space of helium-4 spallation 
processes is open, they are more important than stau-catalyzed 
fusion and hence constrain the stau property.  
\end{abstract}

\maketitle

\section{Introduction}    

Quests for the physics beyond the Standard Model (SM) will reach a new stage 
at the TeV scale.  Among the expected interesting signals of the new physics 
are those provided by exotic charged particles (charged massive particles; 
CHAMPs) with a long lifetime.
The presence of such particles is predicted in many notable models beyond the
SM, although its identity depends on the models one assumes.
CHAMP hunting is indeed one of the major issues of the high energy experiments, 
and its collider phenomenology is enthusiastically studied~\cite{Drees:1990yw, 
Feng:1997zr, Martin:1998vb, Hamaguchi:2004df, Feng:2004yi, Ibarra:2006sz, 
Hamaguchi:2006vu, Feng:2007ke, Kaneko:2008re, Kaneko:2011qi, 
Asai:2009ka,Asai:2011wy,Ito:2010xj,Ito:2009xy}; 
it also motivates other researches including neutrino telescope
observations~\cite{Ahlers:2006pf,Ando:2007ds,Canadas:2008ey} and 
cosmology~\cite{Sigurdson:2003vy, Hisano:2006cj, Borzumati:2008zz, 
Kohri:2009mi}.

Long-lived CHAMPs will play interesting roles in the Big-Bang Nucleosynthesis
(BBN) as well. The light nuclei will interact not only with the CHAMPs during 
the BBN processes~\cite{Pospelov:2008ta, Pospelov:2006sc, 
Kusakabe:2010cb, Kawasaki:2010yh, Kamimura:2008fx, Pradler:2007is, 
Jedamzik:2007cp, Kawasaki:2008qe, Kawasaki:2007xb, Kohri:2006cn, 
Kaplinghat:2006qr, Cyburt:2006uv, Jittoh:2007fr, Jittoh:2008eq, 
Jittoh:2010wh, Bird:2007ge, Bailly:2008yy, Bailly:2009pe}, 
but also with the decay products of the CHAMPs in the post-BBN 
era~\cite{Kohri:2008cf, Pospelov:2010cw, Cyburt:2009pg, Kusakabe:2008kf, 
Cumberbatch:2007me, Kawasaki:2004yh, Kawasaki:2004qu, Jedamzik:2004er}.
The standard scenario of the BBN will thus be altered, and so is the abundance
of the light elements at the present time. One can thus constrain the models 
beyond the Standard Model by evaluating their prediction on the light elements 
abundance and comparing it with the current observations. We can then give 
stringent predictions for the forthcoming experiments and observations according 
to these constraints.

The Standard Model extended with supersymmetry (SUSY) is one of the models 
that can accommodate such long-lived CHAMPs.
With the $R$-parity conservation, the lightest SUSY particle
(LSP) is stable and become a cold dark matter.
Interestingly, it can offer a long-lived CHAMPs if the LSP is the bino-like
neutralino $\tilde{\chi}_{1}^{0}$.
Coannihilation mechanism is required to account for the dark matter abundance in
this case~\cite{Griest:1990kh}, where the LSP and the next-lightest SUSY
particle (NLSP) are almost degenerate in mass.
Staus, denoted by $\tilde{\tau}$ and a possible candidate of the NLSP, can 
acquire a long lifetime when the mass difference with the LSP is less than the 
mass of tau leptons. This is due to the phase space suppression of the final 
state that necessarily consists of three particles or more.
Noting that such long-lived staus will be copious during the
BBN~\cite{Profumo:2004qt,Jittoh:2005pq}, we have shown 
in~\cite{Jittoh:2007fr, Jittoh:2008eq, Jittoh:2010wh} 
that their presence indeed alters the prediction of the standard BBN and
possibly solve the discrepancy of the lithium abundance in the
Universe through the internal conversion reactions.

In this article, we improve our analyses by including new reactions of
\begin{subequations}
\begin{align}
    (\tilde \tau \hspace{0.2mm} ^4\text{He}) 
    &\to \tilde \chi_1^0 + \nu_\tau + \text{t} + \text{n},
    \label{eq:spal-tn}
    \\
    (\tilde \tau \hspace{0.2mm} ^4\text{He}) 
    &\to \tilde \chi_1^0 + \nu_\tau + \text{d} + \text{n} + \text{n}, 
    \label{eq:spal-dnn}
    \\ 
    (\tilde \tau \hspace{0.2mm} ^4\text{He}) 
    &\to \tilde \chi_1^0 + \nu_\tau + \text{p} + \text{n} + \text{n} + \text{n},
    \label{eq:spal-pnnn}
\end{align}
    \label{spa_1}
\end{subequations}
in which $(\tilde \tau \, ^4\text{He})$ represents a bound state of a stau and
$\mathrm{^{4}He}$ nucleus.
Reaction (\ref{spa_1}) is essentially a spallation of the $\mathrm{^{4}He}$
nucleus, producing a triton t, a deuteron d, and neutrons n.
Presence of such spallation processes has been ignored so far
due to the na\"{i}ve expectation that
the rate of the stau-catalyzed fusion~\cite{Pospelov:2006sc}
\begin{equation}
  (\tilde\tau \, \mathrm{^{4}He}) + \mathrm{d}
  \to
  \tilde \tau + \mathrm{^{6}Li}
  \label{eq:Pospelov}
\end{equation}
is larger than the reaction (\ref{spa_1}).
Indeed, the cross section of Eq.~(\ref{eq:Pospelov}) is much larger than that
of $^4$He + d $\to$ $^6$Li + $\gamma$ by (6 -- 7) orders of
magnitude~\cite{Hamaguchi:2007mp}.

We point out that this expectation is indeed na\"{i}ve; the reaction 
Eq.~({\ref{spa_1}}) is more effective than Eq.~({\ref{eq:Pospelov}) 
as long as the spallation processes are kinematically allowed. 
The former reaction rapidly occurs due to the large overlap of their wave 
functions in a bound state.
On the other hand, the latter proceeds slowly since it requires an 
external deuteron which is sparse at the BBN era.  The overproduction of 
t and d is more problematic than that of $\mathrm{^{6}Li}$.
This puts new constraints on the parameters of the minimal supersymmetric 
standard model (MSSM). Note that there 
is no reaction corresponding to Eq.~({\ref{spa_1}}) in the gravitino LSP 
scenario~\cite{Buchmuller:2007ui}.

The purpose of this work is to understand the impact of $^4$He spallation 
processes on light element abundances.  
In Section~\ref{sec:int}, we analytically calculate its reaction rates, 
and compare its timescale with that of the reaction Eq.~({\ref{eq:Pospelov}}).  
In Section~\ref{numerical}, we calculate all of light element abundances 
including exotic reactions, i.e., the $^4$He spallation processes, 
the stau-catalyzed fusion, and the internal conversion processes. 
We show the MSSM parameter space in which we can reproduce the observed 
abundances of both dark matter and light elements including $^7$Li and $^6$Li.  
Section \ref{sec:sum} is devoted to a summary.

\begin{figure}[t]
  \includegraphics[width=65mm]{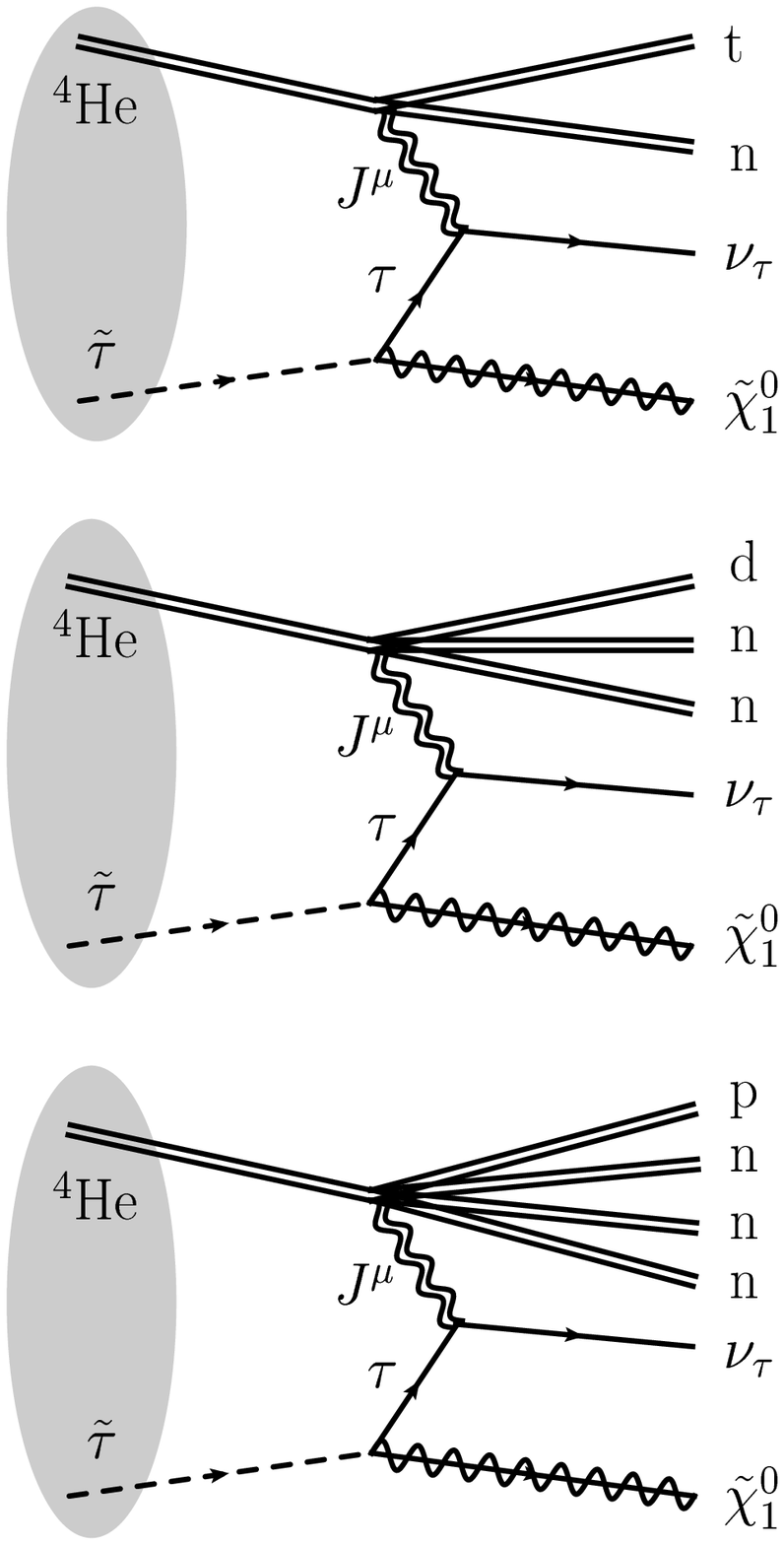}
  \caption{$^4$He spallation processes.}
  \label{fig:spa_diagram}
\end{figure}

\section{Spallation of helium 4}  \label{sec:int} 

Two types of reactions are possible for the bound state of a stau and a
$\mathrm{^{4}He}$ nucleus:
(1) the stau-catalyzed fusion and %
(2) the spallation of the $\mathrm{^{4}He}$ nucleus.
The property of stau is stringently constrained in order to evade the
overproduction of the various light elements due to these processes.

In this section, we calculate the rate of the spallation of the
$\mathrm{^{4}He}$ nucleus.
We compare the result with the rate of the stau-catalyzed fusion to show that
the former is larger than the latter in a large part of the parameter space,
and thereby show that the spallation plays a significant role in the BBN.

The $^4$He spallation processes of Eq.~(\ref{spa_1}) is described by the Lagrangian
\begin{equation}
\begin{split}
    \mathcal{L}
    &
    = \tilde \tau^* \overline{ \tilde \chi_1^0 } 
    (g_\text{L} P_\text{L} + g_\text{R} P_\text{R}) \tau   
    \\[0mm] &
    + \sqrt{2} G_{\textrm{F}}
    \nu_\tau \gamma^\mu P_\text{L} \tau J_\mu    
    + \text{h.c.}, 
\end{split}     \label{Lag}
\end{equation} 
where $G_{\textrm{F}} = 1.166 \times 10^{-5} \mathrm{GeV^{-2}}$ is the 
Fermi coupling constant, $P_{\text{L(R)}}$ represents the chiral projection operator, 
and $J_{\mu}$ is the weak current. The effective coupling constants $g_\text{L}$ 
and $g_\text{R}$ are given by 
\begin{equation}
\begin{split}
    g_\text{L} 
    &= \frac{g}{\sqrt{2} \cos\theta_\text{W}}  
    \sin\theta_\text{W} \cos\theta_\tau ,   \\
    g_\text{R} 
    &= \frac{\sqrt{2} g}{\cos\theta_\text{W}}  
    \sin\theta_\text{W} \sin\theta_\tau 
    \mathrm{e}^{i \gamma_\tau}, 
\end{split}
\end{equation}
where $g$ is the $SU(2)_\text{L}$ gauge coupling constant and
$\theta_{\textrm{W}}$ is the Weinberg angle. The mass eigenstate of staus 
is given by the linear combination of $\tilde \tau_{\textrm{L}}$ and $\tilde 
\tau_{\textrm{R}}$, the superpartners of left-handed and right-handed tau 
leptons, as
\begin{equation}
\begin{split}
    \tilde \tau 
    = \cos\theta_\tau \tilde \tau_\text{L} 
    + \sin\theta_\tau \mathrm{e}^{-i \gamma_\tau} 
    \tilde \tau_\text{R} . 
\end{split}
\end{equation}
Here $\theta_\tau$ is the left-right mixing angle of staus and $\gamma_\tau$ 
is the CP violating phase.

\subsection{$(\tilde \tau \hspace{0.5mm} ^4\text{He}) \to 
\tilde \chi_1^0 + \nu_\tau + \text{t} + \text{n}$ } \label{sec:Tn} 

First we consider the process of Eq.~({\ref{eq:spal-tn}}).
The rate of this process is expressed as 
\begin{equation}
\begin{split}
   \frac{1}{\tau_\text{tn}} 
   = \frac{1}{|\psi|^2 \cdot 
   \sigma v_{\textrm{tn}}}, 
\end{split}     \label{time_Tn}
\end{equation}
where $|\psi|^2$ stands for the overlap of the wave functions of the stau
and the $\mathrm{^4He}$ nucleus.
We estimate the overlap by
\begin{equation}
\begin{split}
   |\psi|^2 = \frac{(Z \alpha m_\text{He})^3}{\pi},
\end{split}     \label{overlap}
\end{equation}
where $Z$ and $m_\text{He}$ represent the atomic number and the mass of $^4$He,
respectively, and $\alpha$ is the fine structure constant.
We assumed that the stau is pointlike particle and is much heavier than
$\mathrm{^{4}He}$ nucleus so that the reduced mass of the bound
state is equal to the mass of $\mathrm{^4He}$ nucleus itself.
The cross section of the elementary process for this reaction is denoted by
$\sigma v_\text{tn}$ and calculated as
\begin{equation}
\begin{split}
   \sigma v_\text{tn}  
   &\equiv 
   \sigma v \bigl( (\tilde \tau ^4\text{He}) \to 
   \tilde \chi_1^0 \nu_\tau \text{tn} \bigr)
   \\ &
   = 
   \frac{1}{2E_{\tilde \tau}} \int
   \frac{d^3 \boldsymbol{p}_\nu}{(2\pi)^3 2E_\nu} 
   \frac{d^3 \boldsymbol{p}_{\tilde \chi}}{(2\pi)^3 2E_{\tilde \chi}} 
   \frac{d^3 \boldsymbol{q}_\text{n}}{(2\pi)^3}
   \frac{d^3 \boldsymbol{q}_\text{t}}{(2\pi)^3} 
   \\ & \hspace{3em} \times 
   \bigl| \mathcal{M} \bigl( (\tilde \tau ^4\text{He}) \to 
   \tilde \chi_1^0 \nu_\tau \text{tn} \bigr) \bigr|^2 
   \\ & \hspace{3em} \times 
   (2 \pi)^4 \delta^{(4)} 
   (p_{\tilde \tau} + p_\text{He} - p_\nu - q_\text{t} - q_\text{n}). 
\end{split}     \label{cross_tn_1}
\end{equation}
Here $p_i$ and $E_i$ are the momentum and the energy of the particle 
species $i$, respectively. 

We briefly show the calculation of the amplitude of this process, leaving the
full calculation in Appendix.
The amplitude is deconstructed as
\begin{equation}
\begin{split}
   &\mathcal{M} \bigl( (\tilde \tau ^4\text{He}) \to 
   \tilde \chi_1^0 \nu_\tau \text{tn} \bigr)
   \\ & \hspace{1em} =
   \langle \text{t \hspace{-1.9mm} n} \hspace{0.5mm} 
   \tilde \chi_1^0 \hspace{0.5mm} \nu_\tau 
   |\mathcal{L}_{\textrm{int}}| 
   ^4\text{He} \hspace{0.5mm} \tilde \tau \rangle 
   \\ & \hspace{1em} = 
   \langle \text{t \hspace{-1.9mm} n} 
   |J^\mu| 
   ^4\text{He} \rangle  ~ 
   \langle \tilde \chi_1^0 \hspace{0.5mm} \nu_\tau 
   |j_\mu| 
   \tilde \tau \rangle.
   \label{ampli_Tn}
\end{split}
\end{equation}
Here we omitted the delta function for the momentum conservation and the spatial integral.
The weak current $J_{\mu}$ consists of a vector current $V_{\mu}$ and an
axial vector current $A_{\mu}$ as $J_{\mu} = V_{\mu} + g_{\textrm{A}} 
A_{\mu}$, where $g_{\textrm{A}}$ is the axial coupling constant.
The relevant components of the currents in this reaction are $V^{0}$ 
and $A^{i}$ ($i = 1, 2, 3$).
We take these operators as a sum of a single-nucleon  operators as
\begin{equation}
  V^{0}
  =
  \sum_{a = 1}^{4} \tau_{a}^{-}
  \mathrm{e}^{\mathrm{i}\boldsymbol{q} \cdot \boldsymbol{r}_{a}}
  \, , \quad
  A^{i}
  =
  \sum_{a = 1}^{4} \tau_{a}^{-} \sigma_{a}^{i}
  \mathrm{e}^{\mathrm{i}\boldsymbol{q} \cdot \boldsymbol{r}_{a}}
  \, ,
\end{equation}
where $\boldsymbol{q}$ is the momentum carried by the current, 
$\boldsymbol{r}_{a}$ is the spatial coordinate of the $a$-th nucleon ($a \in 
\{1, 2, 3, 4 \}$), and $\tau_a^-$ and $\sigma_{a}^{i}$ denote the isospin 
ladder operator and the spin operator of the $a$-th nucleon, respectively. 
Each component leads to a part of hadronic matrix element:
\begin{equation}
\begin{split}
   \langle \text{tn} | V^0 | ^4\text{He} \rangle 
   & = \sqrt{2} \mathcal{M}_{\textrm{tn}} , 
   \\ 
   \langle \text{tn} | g_{\textrm{A}} A^+ | ^4\text{He} \rangle 
   & = \sqrt{2} g_{\textrm{A}} \mathcal{M}_{\textrm{tn}} , 
   \\ 
   \langle \text{tn} | g_{\textrm{A}} A^- | ^4\text{He} \rangle 
   & = - \sqrt{2} g_{\textrm{A}} \mathcal{M}_{\textrm{tn}}, 
   \\
   \langle \text{tn} | g_{\textrm{A}} A^3 | ^4\text{He} \rangle 
   & = - \sqrt{2} g_{\textrm{A}} \mathcal{M}_{\textrm{tn}},
   \label{had_mat} 
\end{split}
\end{equation}
where $A^\pm = (A^1 \pm i A^2)/\sqrt{2}$.
Given the relevant wave functions of a $\mathrm{^4He}$ nucleus, a triton, and a
neutron in Appendix, we obtain the hadronic matrix element as
\begin{equation}
\begin{split}
   &
   \mathcal{M}_{\textrm{tn}} 
   = 
   \biggl(
   \frac{128 \pi}{3} 
   \frac{a_\text{He} a_\text{t}^2}{(a_\text{He} + a_\text{t})^4}
   \biggr)^{3/4} 
   \\ &~~ \times
   \biggl\{ 
   \exp \biggl[ - \frac{\boldsymbol{q}_\text{t}^2}{3 a_\text{He}} \biggr] 
   - \exp \biggl[ - \frac{\boldsymbol{q}_\text{n}^2}{3 a_\text{He}} 
   - \frac{(\boldsymbol{q}_\text{t} + \boldsymbol{q}_\text{n})^2}
   {6 (a_\text{He} + a_\text{t})} \biggr]
   \biggr\}.
   \label{amp_part} 
\end{split}
\end{equation}
Here $\boldsymbol{q}_\text{t}$ and $\boldsymbol{q}_\text{n}$ are three-momenta
of the triton and the neutron, respectively, and $a_\text{He}$ and $a_\text{t}$
are related to the mean square matter radius $R_\text{mat}$ by
\begin{equation}
\begin{split}
   a_\text{He} 
   = \frac{9}{16} \frac{1}{ (R _{\text{mat}})  _{\text{He}} ^2}, ~
   a_\text{t} 
   = \frac{1}{2} \frac{1}{ (R _{\text{mat}}) _{ \text{t}}  ^2}
   \label{a_r} .   
\end{split}
\end{equation}
We list in Table \ref{table:input} input values of the matter radius for the numerical calculation 
in this article.

\begin{table}
\caption{Input values of the matter radius $R_\text{mat}$ 
 for d, t, and $^4\text{He}$, the magnetic radius 
$R_\text{mag}$ for p and n,  nucleus mass $m_X$, excess 
energy $\Delta_X$ for the nucleus $X$, and each reference. 
\label{table:input}}
\begin{center}
\begin{tabular}{llll} \hline
nucleus  
& $R_\text{mat(mag)}$ [fm]/[GeV$^{-1}$] ~
& $m_X$ [GeV] 
& 
$\Delta_X$ [GeV] 
\\[0.7mm] \hline
p
& 0.876 / 4.439 \cite{Borisyuk:2009mg}
& 0.9383 \ \cite{P.Mohr}
& $6.778 \times 10^{-3}$ \ \cite{TOI}
\\[0.7mm] \hline
n
& 0.873 / 4.424 \cite{Kubon:2001rj}
& 0.9396 \ \cite{P.Mohr}
& $8.071 \times 10^{-3}$ \ \cite{TOI}
\\[0.7mm] \hline
d          
& 1.966 / 9.962 \ \cite{Wong:1994sy}
& 1.876 \ \cite{TOI}  ~
& $1.314 \times 10 ^{-2}$ \ \cite{TOI}
\\[0.7mm] \hline
t           
& 1.928 / 9.770 \ \cite{Yoshitake}  
& 2.809 \ \cite{TOI}  
& $1.495 \times 10 ^{-2}$ \ \cite{TOI}
\\[0.7mm] \hline
$^4$He 
& 1.49 / 7.55 \ \cite{Egelhof2001307}
& 3.728  \ \cite{TOI}
& $2.425 \times 10 ^{-3}$ \ \cite{TOI}
\\[0.7mm] \hline
\end{tabular}
\end{center}
\end{table}

The remaining part is straightforwardly calculated to be
\begin{equation}
\begin{split}
   &
   | \langle \tilde \chi_1^0 \hspace{0.5mm} \nu_\tau 
   |j_0| \tilde \tau \rangle |^2
   = 
   | \langle \tilde \chi_1^0 \hspace{0.5mm} \nu_\tau 
   |j_z| \tilde \tau \rangle |^2 
   = 
   4 G_\text{F}^2 |g_\text{R}|^2 
   \frac{m_{\tilde \chi_1^0} E_\nu}{m_\tau^2},
   \\ &
   | \langle \tilde \chi_1^0 \hspace{0.5mm} \nu_\tau 
   |j_\pm| \tilde \tau \rangle |^2 
   = 4 G_\text{F}^2 |g_\text{R}|^2 
   \frac{m_{\tilde \chi_1^0} E_\nu}{m_\tau^2} 
   \biggl( 
   1 \mp \frac{p_\nu^z}{E_\nu}
   \biggr) , 
   \label{leptonic_part}
\end{split} 
\end{equation}
where $E_\nu$ and $p_\nu^z$ are the energy and the $z$-component of the 
momentum of the tau neutrino, respectively. 
We assumed that the stau and the neutralino are non-relativistic.
This equation includes not only all the couplings such as $G_{\textrm{F}}$,
$g_{\textrm{L}}$, and $g_{\textrm{R}}$, but also the effect of the virtual tau
propagation in the Fig.~\ref{fig:spa_diagram}.
Note here that $g_{\textrm{L}}$ coupling does not contribute.
This is because the virtual tau ought to be left-handed at the weak current,
and it flips its chirality during the propagation since the transferred
momentum is much less than its mass.

Combining hadronic part with the other part, we obtain the squared amplitude as
\begin{equation}
\begin{split}
   &
   \bigl| \mathcal{M} \bigl( (\tilde \tau ^4\text{He}) \to 
   \tilde \chi_1^0 \nu_\tau \text{tn} \bigr) \bigr|^2 
   \\ &= 
   \frac{8 m_{\tilde \chi_1 ^0} G_{\text{F}}^2 |g_{\text{R}}|^2}{m_{\tau}^2}
   (1 + 3 g_{A}^2) 
   \mathcal{M}_{\textrm{tn}} ^2  
   E_{\nu} . 
   \label{squ_amp}
\end{split} 
\end{equation}
Integrating on the phase space of the final states, we obtain the cross section as  
\begin{equation}
\begin{split}
   \sigma v_\text{tn} 
   &= 
   \frac{8}{\pi^2} \biggl( \frac{32}{3 \pi} \biggr)^{3/2}
   g^2 \tan^2\theta_W \sin^2\theta_\tau (1+3g_A^2) G_F^2 
   \\ &\times
   \Delta_\text{tn}^4 
   \ \frac{m_\text{t} m_\text{n}}{m_{\tilde \tau} m_\tau^2} \ 
   \frac{a_\text{He}^{3/2} a_\text{t}^3}{(a_\text{He} + a_\text{t})^5} 
   ~I_\text{tn} , 
   \label{cross_tn_2} 
\end{split} 
\end{equation}
\begin{widetext}
\begin{equation}
\begin{split} 
   &
   I_\text{tn} 
   = 
   12 \int_0^1 ds \int_0^{\sqrt{1-s^2}} dt 
   ~\bigl( 1 - s^2 - t^2 \bigr)^2  st
   \\ &\times
   \biggl\{
   \frac{1}{6} \frac{k_\text{t} k_\text{n}}{a_\text{He} + a_\text{t}} 
   s t ~ \exp 
   \biggl[ - \frac{2}{3} \frac{k_\text{t}^2}{a_\text{He}} t^2 \biggr] 
   +
   \frac{1}{4}  \exp 
   \biggl[ - \frac{2}{3} \frac{k_\text{n}^2 }{a_\text{He}} s^2 
   - \frac{1}{3} \frac{k_\text{n}^2 s^2 + k_\text{t}^2 t^2}
   {a_\text{He} + a_\text{t}}  \biggr] 
   \sinh\biggl[ \frac{2}{3} 
   \frac{k_\text{t} k_\text{n}}{a_\text{He} + a_\text{t}} st \biggr]
   \\ & \hspace{2em}
   - 
   \exp
   \biggl[ - \frac{1}{3} \frac{k_\text{n}^2 s^2 + k_\text{t}^2 t^2}{a_\text{He}}  
   - \frac{1}{6} \frac{k_\text{n}^2 s^2 + k_\text{t}^2 t^2}
   {a_\text{He} + a_\text{t}}  \biggr] 
   \sinh\biggl[ \frac{1}{3} 
   \frac{k_\text{t} k_\text{n}}{a_\text{He} + a_\text{t}} st \biggr]
   \biggr\} .
   \label{I_tn} 
\end{split} 
\end{equation}
\end{widetext}
Here $\Delta_\text{tn}$, $k_\text{t}$, and $k_\text{n}$ are defined as 
\begin{equation}
\begin{split}
   &
   \Delta_\text{tn} \equiv \delta m + \Delta_\text{He} - \Delta_\text{t} 
   - \Delta_\text{n} - E_\text{b} ,
   \\ &
   k_\text{t} \equiv \sqrt{2 m _{\text{t}} \Delta_\text{tn} } \ , 
   \\ &
   k_\text{n} \equiv \sqrt{2 m _{\text{n}} \Delta_\text{tn} } \ ,
   \label{k_tn} 
\end{split} 
\end{equation}
where $\Delta_X$ is the excess energy of the nucleus $X$, and $E_\text{b}$ 
is the binding energy of $(\tilde \tau \hspace{0.3mm} ^4\text{He})$ system. 
\subsection{$(\tilde \tau \hspace{0.5mm} ^4\text{He}) \to 
\tilde \chi_1^0 + \nu_\tau + \text{d} + \text{n} + \text{n}$} \label{sec:dnn} 

The rate of another spallation process of Eq.(\ref{eq:spal-dnn}) is similarly 
calculated. The cross section is calculated to be 
%
\begin{widetext}
\vspace{-5mm}
\begin{equation}
\begin{split}
   \sigma v_\text{dnn} 
   &=  
   \frac{192}{\pi^{4}} g^{2} \tan^{2}\theta_{\textrm{W}}
   \sin^{2} \theta_{\tau} G_{\textrm{F}}^{2} \Delta_\text{dnn}^{4}
   \frac{m_{\mathrm{n}}m_{\mathrm{d}}}{m_{\tilde{\tau}} m_{\tau}^{2}}
   \biggl(
   \frac{2a_{\mathrm{d}}}{a_{\mathrm{He}}(a_{\mathrm{d}} + a_{\mathrm{He}})^{2}}
   \biggr)^{3/2}
   I _\text{dnn}
   \, ,
   \label{cross_dnn_2} 
\end{split} 
\end{equation}
where
\begin{equation}
\begin{split}
   &
   I_\text{dnn}
   = 
   \int_0^1 ds \int_0^{\sqrt{1-s^2}} dt \int_0^{\sqrt{1-s^2-t^2}} du 
   (1 - s^2 - t^2 - u^2)^2 
   \\ & \times
   \biggl\{
   (1 + 3g_A^2) a_\text{He} k_\text{n}^3 s t^2 u 
   \exp \biggl[ - \frac{3 k_\text{d}^2 s^2 + 4 k_\text{n}^2 u^2}
   {4 a_\text{He}} \biggr] 
   \sinh \biggl[ \frac{k_\text{n} k_\text{d}}{a_\text{He}} su \biggr] 
   \\ &- 
   \sqrt{2} (1 + g_A^2) a_\text{He} k_\text{n}^3 s t u^2 
   \exp \biggl[ - \frac{3 k_\text{d}^2 s^2 + 2 k_\text{n}^2 t^2 + 
   2 k_\text{n}^2 u^2}{4 a_\text{He}} \biggr] 
   \sinh \biggl[ \frac{1}{\sqrt{2}} \frac{k_\text{n} k_\text{d}}
   {a_\text{He}} st \biggr]
   \\ & +2 \sqrt{2} g_A^2 (a_\text{He} + a_\text{d}) k_\text{n}^3 s t u^2 
   \exp \biggl[ - \frac{k_\text{n}^2 (2 t^2 + u^2)}{2 a_\text{He}} 
   - \frac{k_\text{d}^2 s^2 + 2 k_\text{n}^2 t^2}
   {4 (a_\text{He} + a_\text{d})}\biggr] 
   \sinh \biggl[ \frac{1}{\sqrt{2}} 
   \frac{k_\text{d} k_\text{n}}{a_\text{He} + a_\text{d}} st \biggr] 
   \\ & -16 \sqrt{2} g_A^2 \frac{a_\text{He} ^2}{a_\text{He} + a_\text{d}} 
   \sqrt{5 a_\text{He}^2 +6 a_\text{He} a_\text{d} +2 a_\text{d}^2} 
   k_\text{n} su 
   \\ & \hspace{2em} \times
   \exp (- A_1 k_\text{d}^2 s^2 - A_2 k_\text{n}^2 t^2 
   - A_3 k_\text{n}^2 u^2)
   \sinh (A_4 k_\text{d} k_\text{n} st)
   \sinh (A_5 k_\text{n}^2 tu)
   \biggr\} 
   \ .
   \label{I_dnn} 
\end{split} 
\end{equation}
\end{widetext}
Here $\Delta_\text{dnn}$, $k_\text{d}$, $k_\text{n}$ and $A_i (i=1 \text{ - } 5)$ 
are defined as follows:
\begin{equation}
\begin{split}
   &
   \Delta_\text{dnn} \equiv \delta m + \Delta_\text{He} 
   - \Delta_\text{d} - 2 \Delta_\text{n} - E_b, 
   \\&
   k_\text{n} \equiv \sqrt{2 m _{\text{n}} \Delta_\text{dnn} } \ ,
   \\&
   k_\text{d} \equiv \sqrt{2 m _{\text{d}} \Delta_\text{dnn} } \ ,
   \\&
   A_1 \equiv 
   \frac{4 a_\text{He} + 3a_\text{d}}
   {8 a_\text{He}(a_\text{He} + a_\text{d})} \ ,
   \\&
   A_2 \equiv 
   \frac{22 a_\text{He}^3 + 44 a_\text{He}^2 a_\text{d} 
   + 30 a_\text{He} a_\text{d}^2 + 7 a_\text{d}^3}
   {4 a_\text{He} (a_\text{He} + a_\text{d}) 
   (5 a_\text{He}^2 + 6 a_\text{He} a_\text{d} +2 a_\text{d}^2)} \ ,
   \\&
   A_3 \equiv 
   \frac{8 a_\text{He}^2 + 9 a_\text{He} a_\text{d} +3 a_\text{d}^2}
   {4 a_\text{He} (5 a_\text{He}^2 + 6 a_\text{He} a_\text{d} +2 a_\text{d}^2)} \ ,
   \\&
   A_4 \equiv 
   \frac{1}{4 a_\text{He} (a_\text{He} + a_\text{d})} 
   \sqrt{10 a_\text{He}^2 + 12 a_\text{He} a_\text{d} + 4 a_\text{d}^2} \ ,
   \\&
   A_5 \equiv 
   \frac{(a_\text{He} + a_\text{d})^2}
   {2 a_\text{He} (5 a_\text{He}^2 + 6 a_\text{He} a_\text{d} +2 a_\text{d}^2)} \ .
\end{split} 
\end{equation}
The rate is then obtained in the same manner as Eq.~(\ref{time_Tn}).

\subsection{$(\tilde \tau \hspace{0.5mm} ^4\text{He}) \to 
\tilde \chi_1^0 + \nu_\tau + \text{p} + \text{n} + \text{n} 
+ \text{n}$} \label{sec:pnnn} 

The cross section of spallation process of Eq. (\ref{eq:spal-pnnn}) is calculated 
to be 
\begin{equation}
\begin{split}
   \sigma v_\text{pnnn} 
   &= 
   \frac{8}{\pi^9} 
   \biggl( \frac{32 \pi^3}{a_\text{He}^3} \biggr)^{3/2}
   g^2 \tan^2\theta_\text{W} \sin^2\theta_\tau 
   (1 + 3g_A^2) G_F^2    \\
   &\times
   a_\text{He}  
   \Delta_\text{pnnn}^7 
   \frac{m_\text{N}^5}{m_{\tilde \tau} m_\tau^2} 
   \hspace{0.5mm} I_\text{pnnn} , 
\end{split}
\end{equation}
\begin{widetext}
where
\begin{equation}
\begin{split}
   &
   I _{\text{pnnn}} 
   = 
   \int_0^1 ds    
   \int_0^{\sqrt{1-s^2}} dt 
   \int_0^{\sqrt{1-s^2-t^2}} du    
   \int_0^{\sqrt{1-s^2-t^2-u^2}} dv 
   (1 - s ^2 - t ^2 - u ^2 - v ^2 ) ^2
   s t ^2 u v ^2   
   \\& ~~ \times 
   \biggl\{  
   \frac{1}{\sqrt{2}} \hspace{0.5mm} 
   \exp\biggl[ -\frac{k_N^2}{2 a_\text{He}} 
   (3 s^2 + t^2 + 2 u^2) \biggr]
   \sinh\biggl[ \frac{\sqrt{2} k_N^2}{a_\text{He}} su \biggr] 
    -  
   \exp\biggl[ -\frac{k_N^2}{2 a_\text{He}} 
   \bigl( 3s^2 + t^2 + u^2 + v^2 \bigr) \biggr]
   \sinh\biggl[ \frac{k_\text{N}^2}{a_\text{He}} s u \biggr] 
   \biggr\} , 
\end{split}
\end{equation}
\end{widetext}
where $\Delta_\text{pnnn}$ and $k_\text{N}$ are defined as follows: 
\begin{equation}
\begin{split}
   &
   \Delta_\text{pnnn} \equiv \delta m + \Delta_\text{He} 
   - \Delta_\text{p} - 3 \Delta_\text{n} - E_b , 
   \\ &
   k_\text{N} \equiv \sqrt{2 m_\text{N} \Delta_\text{pnnn} }. 
\end{split} 
\end{equation}
In this calculation, we assumed proton and neutron have an identical kinetic 
energy, and then the factor $k_\text{p}$ and $k_\text{n}$, which are 
introduced to factorize their kinetic energies, are also identical. $k_\text{N}$ 
is the identical factor, and here we took $m_\text{N} = m_\text{n}$.

The reaction rate is obtained in the same manner as Eq.~(\ref{time_Tn}).

\subsection{Comparing the rate of spallation reaction with that of stau-catalyzed fusion}   
\label{sec:rate}   

\begin{figure}[h!]
\includegraphics[width=85mm]{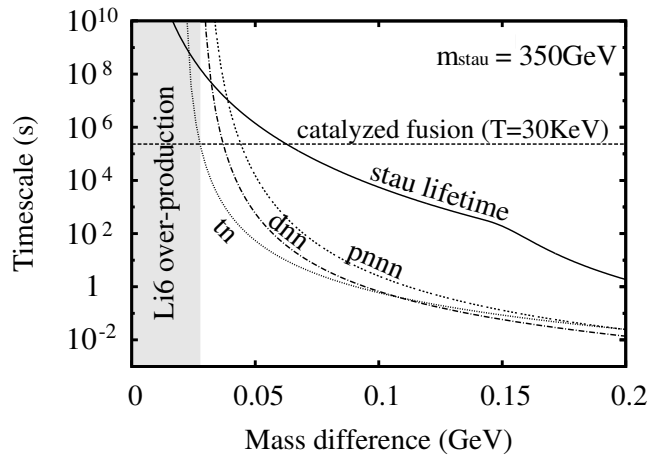}
\caption{Timescale of spallation processes as a function 
of $\delta m$ and the stau-catalyzed fusion at the 
universe temperature $T=30$keV~\cite{Hamaguchi:2007mp}. 
The lifetime of free $\tilde \tau$ (solid line) is also depicted.  
Here we took $m_{\tilde \tau} = 350$GeV, $\sin\theta_\tau 
= 0.8$, and $\gamma_\tau = 0$.}
\label{fig:time}
\end{figure}

We compare the rate of the spallation and that of the stau-catalyzed fusion. 
We first note that the rate of stau-catalyzed fusion strongly depends on the
temperature~\cite{Hamaguchi:2007mp}, and we fix the reference 
temperature to be $30\mathrm{keV}$.
Staus begin to form a bound state with $^4$He at this temperature, which 
corresponds to cosmic time of $10^{3}\mathrm{s}$. Thus the bound 
state is formed when the lifetime of staus is longer than $10^{3}\mathrm{s}$.

Figure~\ref{fig:time} shows the timescale of the spallation processes as a 
function of $\delta m$. The lifetime of free stau is plotted by a solid line. 
We took the reference values of $m_{\tilde \tau} = 350$GeV, 
$\sin\theta_\tau = 0.8$, and $\gamma_\tau = 0$. The inverted rate of 
the stau-catalyzed fusion at the temperature of $30\mathrm{keV}$ is also 
shown by the horizontal dashed line.
Once a bound state is formed, as long as the phase space of spallation 
processes are open sufficiently that is $\delta m \gtrsim 0.026$GeV, 
those processes dominate over other processes.  There $\tilde \tau$ 
property is constrained to evade the over-production of d and/or t. 
For $\delta m \lesssim 0.026$GeV, the dominant process of $(\tilde \tau 
\hspace{0.3mm} ^4\text{He})$ is stau-catalyzed fusion, since the 
free $\tilde \tau$ lifetime is longer than the timescale of stau-catalyzed 
fusion. Thus light gray region is forbidden due to the over-production 
of $^6$Li.

This interpretation of Fig.2 is not much altered by varying the parameters 
relevant with $\tilde \tau$.  First cross sections of spallation processes are 
inversely proportional to $m_{\tilde \tau}$,  and then the timescale of each 
process linearly increases as $m_{\tilde \tau}$ increases. Thus, even when 
$m_{\tilde \tau}$ is larger than $m_{\tilde \tau} = 350$GeV by up to a 
factor of ten, the region of $^6$Li over-production scarcely changes.  
Next we point out that our result depend only mildly on the left-right mixing 
of the stau.  Indeed, cross section of the $^4 \text{He}$ spallation is 
proportional to $\sin ^2 \theta _{\tau}$.  Its order of magnitude will not 
change as long as the right-handed component is significant.

\section{Light elements abundances and allowed  parameter space}  
\label{numerical}   

We numerically calculate the primordial abundances of light elements
including  $^4$He spallation processes and $\tilde \tau$ catalyzed
nuclear fusion. Then we can  search for  allowed regions of the
parameter space to fit observational light element abundances.

So far it has been reported that there is a discrepancy between the
theoretical value of $^7$Li abundance predicted in the standard BBN
(SBBN)  and the observational one. This is called $^7$Li problem.
SBBN predicts the $^7$Li to H ratio to be ${\rm
Log}_{10}(^7\text{Li}/\text{H}) = -9.35 \pm 0.06$ when we adopt
a recent value of   baryon to photon ratio $\eta = (6.225 \pm 0.170)
\times 10^{-10}$ (68\% C.L.)  reported by the WMAP
satellite~\cite{Komatsu:2010fb}, and  experimental data of the rate
for the $^{7}$Li or $^{7}$Be production through
$^3$He~+~$^4$He~$\to~^7$Be~+~$\gamma$~\cite{Cyburt:2008kw} 
($^{7}$Li is produced from $^7$Be by its electron capture, $^7$Be~+~$e^-$
$\to$  $^{7}$Li + $\nu_e$ at a later epoch).
On the other hand, the primordial $^7$Li abundance is observed in
metal-poor halo stars as absorption lines~\cite{Spite:1982dd}.  Recent
observationally-inferred value of the primordial $^7$Li to hydrogen
ratio is ${\rm Log}_{10}(^7\text{Li}/\text{H}) = -9.63 \pm
0.06$~\cite{Melendez:2004ni} for a high value, and ${\rm
Log}_{10}(^7\text{Li}/\text{H}) = -9.90 \pm
0.09$~\cite{Bonifacio:2006au} for a low value. (See also
Refs.~\cite{Ryan:2000zz, Asplund:2005yt,  Aoki:2009ce} for another
values.)
Therefore there is a discrepancy at more than three sigma
between theoretical and observational values even when we adopt the
high value of \cite{Melendez:2004ni}.   This discrepancy can be hardly
attributed to the correction of the cross section of nuclear
reaction~\cite{Cyburt:2003ae,Angulo:2005mi}.  Even if we consider
nonstandard astrophysical models such as those including diffusion
effects~\cite{Richard:2004pj,Korn:2006tv}, it might be difficult to
fit all of the data consistently~\cite{Lind:2009ta}.

In Figs.~\ref{fig:y_deltamMR} and \ref{fig:y_deltamBoni}, we plot the
allowed parameter regions which are obtained by comparing the
theoretical values to observational ones for the high and low
$^7$Li/H, respectively. 
Vertical axis is the yield value of $\tilde \tau$ at the time of the formation 
of the bound states with nuclei, $Y_{\tilde \tau} = n_{\tilde \tau}/s$ 
($s$ is the entropy density), and horizontal axis is the mass difference of 
$\tilde \tau$ and $\tilde \chi_1^0$. 
We have adopted
following another observational constraints on the light element
abundances: an upper bound on the $^6$Li to $^7$Li ratio,
$^6$Li/$^7$Li $< 0.046 + 0.022$~\cite{Asplund:2005yt}, the deuteron to
hydrogen ratio, D/H=$(2.80 \pm 0.20) \times
10^{-5}$~\cite{Pettini:2008mq}, and an upper bound on the $^3$He to
deuteron ratio, $^3$He/D $< 0.87 + 0.27$~\cite{GG03}.

The solid line (orange line) denotes a theoretical value of the  thermal relic
abundance for staus~\cite{Jittoh:2010wh} while keeping
observationally-allowed dark matter density  $\Omega_{\rm DM}h^{2} =
0.11 \pm 0.01$ (2 $\sigma$)~\cite{Komatsu:2010fb}  as  total
$\chi_{1}^{0}+\tilde{\tau}$ abundance. For reference, we also plot the
observationally-allowed dark matter density in the figures by a
horizontal band.

At around $\delta m \sim 0.1$ GeV, we find that $^7$Li/H can be fitted
to the observational value without conflicting with the other light element
abundances.~\footnote{See also~\cite{Jedamzik:2004er, Bird:2007ge, 
Pospelov:2010cw, Kawasaki:2010yh, Erken:2011vv} 
for another mechanisms to reduce
$^7$Li/H. } As shown in Fig.~\ref{fig:y_deltamMR},  it should be
impressive that the relic density is consistent with the  allowed
region at $Y_{\tilde{\tau}} = 2 \times 10^{-13}$ at 3 $\sigma$ in case
of the high value of $^7$Li/H in~\cite{Melendez:2004ni}.

\begin{figure}[t!]
    \includegraphics[width=85mm]{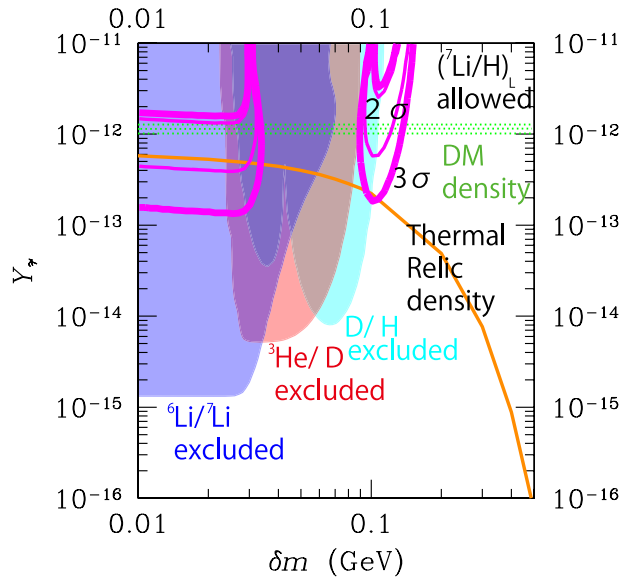}
    \caption{
    Allowed regions from observational light element abundances at 2
    $\sigma$.  Here we have adopted the higher value of the
    observational  $^7$Li/H in \cite{Melendez:2004ni} denoted by
    $(^7$Li/H)$_{\rm H}$, and have plotted both the 2$\sigma$ (thin
    line) and 3$\sigma$ (thick line ) only for $^7$Li/H. The
    horizontal band means the observationally-allowed dark matter
    density. We have adopted $m_{\tilde{\tau}} = 350$~GeV,
    $\sin\theta_{\tau} = 0.8$, and $\gamma_{\tau}=0$, respectively.  }
    \label{fig:y_deltamMR}
\end{figure}

\begin{figure}[t!]
    \includegraphics[width=85mm]{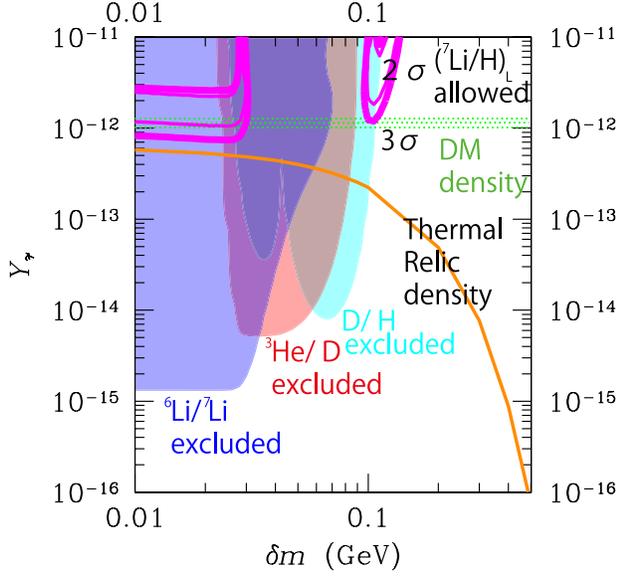}
    \caption{Same as Fig.~\ref{fig:y_deltamMR}, but for the lower
    value of observational 
     $^7$Li/H reported in
    \cite{Bonifacio:2006au}, which is denoted by $(^7$Li/H)$_{\rm L}$.}
    \label{fig:y_deltamBoni}
\end{figure}

\section{Summary}  \label{sec:sum} 

We calculated primordial abundances of all of light elements involving 
the helium-4 spallation processes, the catalyzed fusion, and the internal 
conversion processes.
Newly included in the present work is the spallation of the $^4$He 
in the stau-$^4$He bound state given in Eq.(1). 
This process is only present in the model which predicts the long-lived 
charged particles due to the phase space suppression with 
the weakly interacting daughter particle.

We calculated the rate of the helium-4 spallation processes 
analytically, and compared it with that of catalyzed fusion.   
We found that the spallation of $^4$He nuclei dominate over the 
catalyzed fusion as long as the phase space of the spallation 
processes are open and hence the property of long lived stau 
is constrained from avoiding the overproduction of a deuteron 
and/or a triton. 
In spite of these new constraints, we found that the lithium discrepancy 
and the dark matter abundance can be simultaneously solved in the 
parameter regions presented in Fig.~\ref{fig:y_deltamMR}.

\section*{Acknowledgments}   

We are very grateful to Masayasu Kamimura for helpful discussions 
on the nuclear reactions. 
The work of K.S. was financially supported by the Sasakawa Scientific
Research  Grant from The Japan Science Society.  This  work was
supported in part by the Grant-in-Aid for the Ministry of Education,
Culture, Sports, Science, and Technology, Government of Japan,
No. 18071001 (K.K.), No. 22244030 (K.K.), No. 21111006 (K.K.), 
No. 22740140 (M.K.), No. 20540251 (J.S.). and No. 23740208 (M.Y.).
K.K. was also partly supported by the Center for the Promotion of
Integrated Sciences (CPIS) of Sokendai.

\appendix 
\section{Amplitude of the Spallation Reaction of $^4$He} 

The interaction relevant to our scenario is
\begin{equation}
\begin{split}
  \mathcal{L}_{\textrm{int}}
  = &
  \tilde{\tau}^{\ast} \overline{\tilde{\chi}^{0} _{1}}
  (g_{\textrm{L}} P_{\textrm{L}} + g_{\textrm{R}} P_{\textrm{R}})
  \tau
  \\ &
  +
  \sqrt{2}G_{\textrm{F}}
  (\nu_{\tau} \gamma^{\mu} P_{\textrm{L}} \tau) J_{\mu}
  + \textrm{H.c.}
  \, ,
\end{split}
\end{equation}
where $J_{\mu}$ is the hadronic current.
This interaction allows the spallation of the nuclei such as
\begin{subequations}
\begin{align}
    (\tilde \tau \hspace{0.2mm} ^4\text{He}) 
    &\to \tilde \chi_1^0 + \nu_\tau + \text{t} + \text{n},
    \label{eq:spallation-tn}
    \\
    (\tilde \tau \hspace{0.2mm} ^4\text{He}) 
    &\to \tilde \chi_1^0 + \nu_\tau + \text{d} + \text{n} + \text{n}, 
    \label{eq:spallation-dnn}
    \\
    (\tilde \tau \hspace{0.2mm} ^4\text{He}) 
    &\to \tilde \chi_1^0 + \nu_\tau + \text{p} + \text{n} + \text{n} + \text{n}, 
    \label{eq:spallation-pnnn}
\end{align}  
    \label{spallation_1}
\end{subequations}
\hspace{-1.2mm}to take place. 
In this section, we illustrate how we calculate the amplitude of the processes
of Eqs.~(\ref{spallation_1}).

\subsection{$^4\text{He} \to \text{tn}$}

The amplitude of the process (\ref{eq:spallation-tn}) is given by
\begin{equation}
\begin{split}
   &\mathcal{M} \bigl( (\tilde \tau ^4\text{He}) \to 
   \tilde \chi_1^0 \nu_\tau \text{tn} \bigr)
   \\ & \hspace{1em} = 
   \langle \text{t \hspace{-1.9mm} n} 
   |J^\mu| 
   ^4\text{He} \rangle  ~ 
   \langle \tilde \chi_1^0 \hspace{0.5mm} \nu_\tau 
   |j_\mu| 
   \tilde \tau \rangle.
   \label{eq:amp-def}
\end{split}
\end{equation}
in which we define the leptonic matrix element by
\begin{equation}
\begin{split}
  &
  \langle \tilde{\chi}^{0} _{1} \, \nu_{\tau} \vert
    j_{\mu}
  \vert \tilde{\tau} \rangle
  \\ &
  =
  \sqrt{2} G_{\textrm{F}}
  \langle \tilde{\chi}^{0} _{1} \, \nu_{\tau} \vert
  [\nu_{\tau} \gamma_{\mu} P_{\textrm{L}} \tau]
  [\bar{\tau}
   (g_{\textrm{L}} P_{\textrm{R}} + g_{\textrm{R}} ^{\ast} P_{\textrm{L}})
  \tilde{\chi}^{0} _{1} \tilde{\tau}]
  \vert \tilde{\tau} \rangle
\end{split}
\end{equation}
and the hadronic matrix element by 
\begin{math}
  \langle \mathrm{t} \, \mathrm{n} \vert
    J^{\mu}
  \vert \mathrm{^{4}He} \rangle
  \, .
\end{math}
We separately calculate these two matrix elements.

\subsubsection{Leptonic matrix element}  

The leptonic matrix element is directly calculated under
the following simplifications:
\begin{itemize}
  \item Neutralino is treated as non-relativistic particle
since its mass $m _{\tilde \chi_1^0}$ is much larger than the mass
of nuclei.
  \item The momentum of a virtual tau is negligibly smaller
than its mass due to the assumption that the stau and
the neutralino are nearly degenerate with the mass difference
of $O(10 - 100)$MeV.
\end{itemize}
A straightforward calculation leads to
\begin{equation}
\begin{split}
   &
   | \langle \tilde \chi_1^0 \hspace{0.5mm} \nu_\tau 
   |j_0| \tilde \tau \rangle |^2
   = 
   | \langle \tilde \chi_1^0 \hspace{0.5mm} \nu_\tau 
   |j_3| \tilde \tau \rangle |^2 
   = 
   4 G_\text{F}^2 |g_\text{R}|^2 
   \frac{m_{\tilde \chi_1^0} E_\nu}{m_\tau^2},
   \\ &
   | \langle \tilde \chi_1^0 \hspace{0.5mm} \nu_\tau 
   |j_\pm| \tilde \tau \rangle |^2 
   = 4 G_\text{F}^2 |g_\text{R}|^2 
   \frac{m_{\tilde \chi_1^0} E_\nu}{m_\tau^2} 
   \biggl( 
   1 \mp \frac{p_\nu^z}{E_\nu}
   \biggr) , 
   \label{leptonic_part_Ap}
\end{split} 
\end{equation}
where $E_i , p_i$, and $m_i$ individually stand for energy, four-momentum and 
mass of particles.

\subsubsection{Hadronic matrix element}  

Calculation of the hadronic matrix element requires the explicit form of the
hadronic current and the wave functions of the nuclei.

We need the wave functions of initial helium, final triton, and nucleon.
Building up these wave functions requires special attention to the symmetry.
The wave function consists of spatial, spin, and isospin parts, and should be
antisymmetric under the exchange of the two nucleons.
The spin and isospin of the nucleus dictates the spin and isospin part of the
wave function. We then arrange the spatial part so that the total wave function 
be antisymmetric under the permutation of the nucleons. We model the spatial 
wave functions by Gaussian functions in terms of Jacobi coordinates.

Let us make a wave function of $\mathrm{^{4}He}$ by this prescription.
The spin and isospin parts of the wave function is constructed according to $S =
0$ and $I = 0$, and turns out to be
\begin{equation}
\begin{split}
  &
  \bigl\vert \mathrm{^{4}He} \bigr\rangle
  =
  \\ &
  \frac{1}{2\sqrt6} \Bigl[
    \vert \mathrm{pnpn} \rangle
    \bigl(
    \vert {\uparrow \uparrow \downarrow \downarrow} \rangle
    + \vert {\downarrow \downarrow \uparrow \uparrow} \rangle
    - \vert {\uparrow \downarrow \downarrow \uparrow} \rangle
    - \vert {\downarrow \uparrow \uparrow \downarrow} \rangle
    \bigr)
    \\ &
    + \vert \mathrm{pnnp} \rangle
    \bigl(
      - \vert {\uparrow \uparrow \downarrow \downarrow} \rangle
      - \vert {\downarrow \downarrow \uparrow \uparrow} \rangle
      + \vert {\uparrow \downarrow \uparrow \downarrow} \rangle
      + \vert {\downarrow \uparrow \downarrow \uparrow}\rangle
    \bigr)
    \\ &
    + \vert \mathrm{nppn} \rangle
    \bigl(
      - \vert {\uparrow \uparrow \downarrow \downarrow} \rangle
      - \vert {\downarrow \downarrow \uparrow \uparrow} \rangle
      + \vert {\uparrow \downarrow \uparrow \downarrow} \rangle
      + \vert {\downarrow \uparrow \downarrow \uparrow} \rangle
    \bigr)
    \\ &
    + \vert \mathrm{npnp} \rangle
    \bigl(
      \vert {\uparrow \uparrow \downarrow \downarrow} \rangle
      + \vert {\downarrow \downarrow \uparrow \uparrow} \rangle
      - \vert {\uparrow \downarrow \downarrow \uparrow} \rangle
      - \vert {\downarrow  \uparrow \uparrow \downarrow} \rangle
    \bigr)
    \\ &
    +
    \vert \mathrm{ppnn} \rangle
    \bigl(
    - \vert {\uparrow \downarrow \uparrow \downarrow} \rangle
    + \vert {\uparrow \downarrow \downarrow \uparrow} \rangle
    + \vert {\downarrow \uparrow \uparrow \downarrow} \rangle
    - \vert {\downarrow \uparrow \downarrow \uparrow} \rangle
    \bigr)
    \\ &
    +
    \vert \mathrm{nnpp} \rangle
    \bigl(
    - \vert {\uparrow \downarrow \uparrow \downarrow} \rangle
    + \vert {\uparrow \downarrow \downarrow \uparrow} \rangle
    + \vert {\downarrow \uparrow \uparrow \downarrow} \rangle
    - \vert {\downarrow \uparrow \downarrow \uparrow} \rangle
    \bigr)
  \Bigr]
  \, .
\end{split}
\label{eq:4He-si}
\end{equation}
The above wave function is antisymmetric under the exchange of two particles.
Thus the spatial part ought to be symmetric, and is constructed as
\begin{equation}
\begin{split}
   &
   \psi_{\mathrm{He}}( \boldsymbol{r}_{1}, \boldsymbol{r}_{2},
   \boldsymbol{r}_{3}, \boldsymbol{r}_{4}) 
   = \Bigl( 2 \frac{a_{\mathrm{He}}^{3}}{\pi^{3}} \Bigr)^{3/4}
   \\& \times
   \exp \Bigl\{
   - a_{\mathrm{He}} \Bigl[
     \boldsymbol{r}_{1}^{2}
     + \boldsymbol{r}_{2}^{2}
     + \boldsymbol{r}_{3}^{2}
     + \boldsymbol{r}_{4}^{2}
     \\& ~~ - 
     \frac{1}{4}
     (\boldsymbol{r}_{1} + \boldsymbol{r}_{2}
         + \boldsymbol{r}_{3} + \boldsymbol{r}_{4})^{2}
   \Bigr]
  \Bigr\}
\end{split}  \label{eq:4He-space}
\end{equation}
where $a_{\mathrm{He}}$ parameterizes the nuclear radius and related to mean square 
matter radius $R_\text{mat}$ by 
\begin{equation}
  a_{\mathrm{He}} = \frac{9}{16} \frac{1}{( R_\text{mat} ) _{\mathrm{He}} ^2 }
  \, .
\end{equation}
We applied the non-relativistic normalization for the nuclear wave function,
which is
\begin{multline}
  \int
  \mathrm{d}^{3}\boldsymbol{r}_{1}
  \mathrm{d}^{3}\boldsymbol{r}_{2}
  \mathrm{d}^{3}\boldsymbol{r}_{3}
  \mathrm{d}^{3}\boldsymbol{r}_{4}
  \, 
  \vert \psi_{\mathrm{He}}(\boldsymbol{r}_{1}, \boldsymbol{r}_{2},
  \boldsymbol{r}_{3}, \boldsymbol{r}_{4}) \vert^{2}
  \\ \times
  \delta^{3} \Bigl(
    \frac{1}{4}
    (\boldsymbol{r}_{1} + \boldsymbol{r}_{2}
     + \boldsymbol{r}_{3} + \boldsymbol{r}_{4})
  \Bigr)
  = 1
  \, .
  \label{eq:4He-space-norm}
\end{multline}
The wave function (\ref{eq:4He-space}) is independent of the coordinate of the
center of mass since the initial $\mathrm{^{4}He}$ is taken to be stationary.
The complete wave function of $\mathrm{^{4}He}$ nucleus is a direct product of
Eqs.~(\ref{eq:4He-si}) and (\ref{eq:4He-space}).

The wave functions of other nuclei are similarly obtained.
For the triton, the spin and isospin part of the wave function is formed
according to $S = 1/2, I = 1/2,$ and $I_{z} = -1/2$; it is given by
\begin{subequations}%
\begin{align}%
  &
\begin{split}
  \vert {\mathrm{t} \uparrow} \rangle
  =
  \frac{1}{\sqrt{6}}
  \Bigl[
  &
    \vert \mathrm{pnn} \rangle
    \bigl(
      \vert {\uparrow \uparrow \downarrow} \rangle
      - \vert {\uparrow \downarrow \uparrow} \rangle
    \bigr)
    \\ &
    -
    \vert \mathrm{npn} \rangle
    \bigl(
      \vert {\uparrow \uparrow \downarrow} \rangle
      - \vert {\downarrow \uparrow \uparrow} \rangle
    \bigr)
    \\ &
    +
    \vert \mathrm{nnp} \rangle
    \bigl(
      \vert {\uparrow \downarrow \uparrow} \rangle
      - \vert {\downarrow \uparrow \uparrow} \rangle
    \bigr)
  \Bigr]
  \, ,
\end{split}
  \\ &
\begin{split}
  \vert {\mathrm{t} \downarrow} \rangle
  =
  \frac{1}{\sqrt{6}}
  \Bigl[
  &
    \vert \mathrm{pnn} \rangle
    \bigl(
      \vert {\downarrow \uparrow \downarrow} \rangle
      - \vert {\downarrow \downarrow \uparrow} \rangle
    \bigr)
    \\ &
    -
    \vert \mathrm{npn} \rangle
    \bigl(
      \vert {\uparrow \downarrow \downarrow} \rangle
      - \vert {\downarrow \downarrow \uparrow} \rangle
    \bigr)
    \\ &
    +
    \vert \mathrm{nnp} \rangle
    \bigl(
      \vert {\uparrow \downarrow \downarrow} \rangle
      - \vert {\downarrow \uparrow \downarrow} \rangle
    \bigr)
  \Bigr]
  \, .
\end{split}
\end{align}%
\end{subequations}%
Its spatial wave function is
\begin{equation}
\begin{split}
  &
  \psi_{\mathrm{t}}
  (\boldsymbol{r}_{1}, \boldsymbol{r}_{2}, \boldsymbol{r}_{3})
  \\ &
  =
  \Bigl( \frac{4}{3} \frac{a_{\mathrm{t}}^{2}}{\pi^{2}} \Bigr)^{3/4}
  \exp \Bigl(
    \mathrm{i} \boldsymbol{q}_{\textrm{t}} \cdot
    \frac{1}{3} (\boldsymbol{r}_{1} + \boldsymbol{r}_{2} + \boldsymbol{r}_{3})
  \Bigr)
  \\ & \times
  \exp \Bigl\{
   - a_{\mathrm{t}} \Bigl[
     \boldsymbol{r}_{1}^{2} + \boldsymbol{r}_{2}^{2} + \boldsymbol{r}_{3}^{2}
     -
     \frac{1}{3}
     (\boldsymbol{r}_{1} + \boldsymbol{r}_{2} + \boldsymbol{r}_{3})^{2}
   \Bigr]
  \Bigr\}
  \, ,
\end{split}
\label{eq:t-space}
\end{equation}
where $\boldsymbol{q}_{\textrm{t}}$ is the center-of-mass momentum of the triton
and $a_{\mathrm{t}}$ is related to the mean square matter radius as
\begin{equation}
  a_{\mathrm{t}} = \frac{1}{2} \frac{1}{( R _{\text{mat}} )_{\mathrm{t}} ^2}
  \, .
\end{equation}
The wave function of the final neutron is simply taken to be
\begin{equation}
  \psi_{\mathrm{n}}(\boldsymbol{r}) =
  \mathrm{e}^{\mathrm{i} \boldsymbol{q}_{\mathrm{n}} \cdot \boldsymbol{r}}
\end{equation}
with trivial spin and isospin parts.

Using these wave functions, we can calculate hadronic matrix elements of Eq.~(\ref{eq:amp-def}) as
\begin{equation}
\begin{split}
   \langle \text{tn} | J ^{0} | ^4\text{He} \rangle 
   & = \sqrt{2} \mathcal{M}_{\textrm{tn}} \, , 
   \\ 
   \langle \text{tn} | J ^{+} | ^4\text{He} \rangle 
   & = \sqrt{2} g_{\textrm{A}} \mathcal{M}_{\textrm{tn}} \, ,
   \\ 
   \langle \text{tn} | J ^{-} | ^4\text{He} \rangle 
   & = - \sqrt{2} g_{\textrm{A}} \mathcal{M}_{\textrm{tn}} \, ,
   \\
   \langle \text{tn} | J ^{3} | ^4\text{He} \rangle 
   & = - \sqrt{2} g_{\textrm{A}} \mathcal{M}_{\textrm{tn}} \, ,
   \label{had_mat2} 
\end{split}
\end{equation}
where $\mathcal{M} _{\textrm{tn}}$ is defined as
\begin{equation}
\begin{split}
   &
   \mathcal{M}_{\textrm{tn}} 
   = 
   \biggl(
   \frac{128 \pi}{3} 
   \frac{a_\text{He} a_\text{t}^2}{(a_\text{He} + a_\text{t})^4}
   \biggr)^{3/4} 
   \\ &~~ \times
   \biggl\{ 
   \exp \biggl[ - \frac{\boldsymbol{q}_\text{t}^2}{3 a_\text{He}} \biggr] 
   - \exp \biggl[ - \frac{\boldsymbol{q}_\text{n}^2}{3 a_\text{He}} 
   - \frac{(\boldsymbol{q}_\text{t} + \boldsymbol{q}_\text{n})^2}
   {6 (a_\text{He} + a_\text{t})} \biggr]
   \biggr\} \, .
   \label{amp_part_2} 
\end{split}
\end{equation}
%

\subsubsection{Amplitude} 

Combining the hadronic matrix elements with the leptonic matrix elements, 
we obtain the square amplitude of Eq.~(\ref{eq:amp-def}) as 
\begin{align}
   & | \mathcal{M} \bigl( (\tilde \tau ^4\text{He}) \to 
   \tilde \chi_1^0 \nu_\tau \text{tn} \bigr) |^2 \notag \\
   &= 
   | \langle \text{t \hspace{-1.9mm} n} 
   |J ^0| ^4\text{He} \rangle |^2 ~ 
   | \langle \tilde \chi_1^0 \hspace{0.5mm} \nu_\tau 
   |j _0| \tilde \tau \rangle |^2 \notag \\
   &+ 
   | \langle \text{t \hspace{-1.9mm} n} 
   |J ^-| ^4\text{He} \rangle |^2 ~ 
   | \langle \tilde \chi_1^0 \hspace{0.5mm} \nu_\tau 
   |j _+| \tilde \tau \rangle |^2\notag 
   \\ &+ 
   | \langle \text{t \hspace{-1.9mm} n} 
   |J^+| ^4\text{He} \rangle |^2 ~ 
   | \langle \tilde \chi_1^0 \hspace{0.5mm} \nu_\tau 
   |j _-| \tilde \tau \rangle |^2 \notag \\
   &+ 
   | \langle \text{t \hspace{-1.9mm} n} 
   |J ^3| ^4\text{He} \rangle | ^2 ~ 
   | \langle \tilde \chi_1^0 \hspace{0.5mm} \nu_\tau 
   |j _3| \tilde \tau \rangle |^2 \notag 
   \\&= 
   \frac{8 m_{\tilde \chi_1^0} G_{\text{F}}^2 |g_\text{R}|^2}
   {m_\tau^2} (1 + 3 g_\text{A}^2) \mathcal{M}_\textrm{tn}^2 E_\nu 
\end{align}

\subsection{$^4\text{He} \to \text{dnn}$} 

The calculation presented in the previous section is applicable to 
another spallation process of Eq.~({\ref{eq:spallation-dnn}}).  
The amplitude we need is
\begin{equation}
\begin{split}
  &
  \mathcal{M}(\mathrm{^{4}He} \to \mathrm{dnn})
  \\ &
  =
  \langle \tilde{\chi}^{0} _{1} \, \nu_{\tau} \vert
    j_{\mu}
  \vert \tilde{\tau} \rangle
  \langle \mathrm{d \, n \, n} \vert J^{\mu} \vert \mathrm{^{4}He} \rangle
  \, .
  \label{eq:amp-dnn-def}
\end{split}  
\end{equation}
The leptonic matrix element is same as in Eq.~(\ref{eq:amp-def}) and is already
calculated in Eq.~(\ref{leptonic_part_Ap}), while the hadronic matrix
element requires a calculation anew.

\subsubsection{Hadronic matrix element}  

We need the wave function of the final states, which include a deuteron and two
neutrons.
The spin and isospin of a deuteron is $S = 1$ and $I = 0$, and the corresponding
wave functions are
\begin{align}
 &\vert {\mathrm{d}, +1} \rangle
   =
  \frac{1}{\sqrt{2}}
  \bigl(
    \vert {\mathrm{p} \uparrow} \rangle
    \vert {\mathrm{n} \uparrow} \rangle
    -
    \vert {\mathrm{n} \uparrow} \rangle
    \vert {\mathrm{p} \uparrow} \rangle
  \bigr)
  \, ,
  \\
 &\vert {\mathrm{d}, 0} \rangle
   =
  \frac{1}{2}
  \bigl(
    \vert {\mathrm{p} \uparrow} \rangle
    \vert {\mathrm{n} \downarrow} \rangle
    -
    \vert {\mathrm{n} \uparrow} \rangle
    \vert {\mathrm{p} \downarrow} \rangle
    \nonumber \\ & \hspace{6em}
    +
    \vert {\mathrm{p} \downarrow} \rangle
    \vert {\mathrm{n} \uparrow} \rangle
    -
    \vert {\mathrm{n} \downarrow} \rangle
    \vert {\mathrm{p} \uparrow} \rangle
  \bigr)
  \, ,
  \\
  &\vert {\mathrm{d}, -1} \rangle
    =
  \frac{1}{\sqrt{2}}
  \bigl(
    \vert {\mathrm{p} \downarrow} \rangle
    \vert {\mathrm{n} \downarrow} \rangle
    -
    \vert {\mathrm{n} \downarrow} \rangle
    \vert {\mathrm{p} \downarrow} \rangle
  \bigr) 
    \, .
\end{align}
The spatial part of the wave function is given by
\begin{equation}
\begin{split}
   &
   \psi_\text{d} (\boldsymbol{r}_1, \boldsymbol{r}_2) 
   = 
   \Bigl( \frac{a_\text{d}}{\pi} \Bigr)^{3/4} 
   \\& \times 
   \exp \Bigl(
   \text{i} \boldsymbol{q}_\text{d} \cdot
   \frac{\boldsymbol{r}_1 + \boldsymbol{r}_2}{2}
   \Bigr)
   \exp \Bigl[
   - \frac{1}{2} a_\text{d} (\boldsymbol{r}_1 - \boldsymbol{r}_2)^{2}
   \Bigr] \ ,
\end{split}  
\end{equation}
where $\boldsymbol{q}_{\mathrm{d}}$ is the center-of-mass momentum %
and $a_{\mathrm{d}}$ is related to the mean square matter radius as
\begin{equation}
  a_{\mathrm{d}} = \frac{3}{8} \frac{1}{( R _{\text{mat}} ) _{\mathrm{d}} ^2}
  \, .
\end{equation}
The spin of two neutrons can be $S = 0$ and $S = 1$. 
For each case, spin and isospin parts of the wave functions are
\begin{align}
 & \vert {\mathrm{n} _{0}} \rangle
   =
   \frac{1}{\sqrt{2}}
  \bigl(
    \vert {\mathrm{n} \uparrow} \rangle
    \vert {\mathrm{n} \downarrow} \rangle
    -
    \vert {\mathrm{n} \downarrow} \rangle
    \vert {\mathrm{n} \uparrow} \rangle \, , \\
 &\vert {\mathrm{n} _{1}, +1} \rangle
   =
    \vert {\mathrm{n} \uparrow} \rangle
    \vert {\mathrm{n} \uparrow} \rangle \, , \\
 &\vert {\mathrm{n} _{1}, 0} \rangle
   =
  \frac{1}{\sqrt{2}}
  \bigl(
    \vert {\mathrm{n} \uparrow} \rangle
    \vert {\mathrm{n} \downarrow} \rangle
    +
    \vert {\mathrm{n} \downarrow} \rangle
    \vert {\mathrm{n} \uparrow} \rangle \, , \\
 &\vert {\mathrm{n} _{1}, -1} \rangle
    =
    \vert {\mathrm{n} \downarrow} \rangle
    \vert {\mathrm{n} \downarrow} \rangle \, , 
\end{align}
where $| \text{n} _{i} \rangle $ expresses spin and isospin part of wave function of $S = i$.  
Since the spin and isospin part of $S = 1$ are symmetric under the exchange of two particles, 
the spatial part of the wave function ought to be antisymmetric.
The spin and isospin part of $S = 0$, on the other hand, is antisymmetric under the exchange 
of two particles, then the spatial part of the wave function ought to be symmetric.
Therefore the spatial parts of each wave function are given by 
\begin{align}
  \psi _{\mathrm{n} 0} (\boldsymbol{r}_{1}, \boldsymbol{r}_{2})
  &=
  \frac{1}{\sqrt{2}} 
     \{ 
      \exp \left [ i (\boldsymbol{q} _{\mathrm{n}1} \cdot \boldsymbol{r} _{1} +
                           \boldsymbol{q} _{\mathrm{n}2} \cdot \boldsymbol{r} _{2} ) \right ] \notag \\  
  &\hspace{12mm}
   + \exp \left [ i (\boldsymbol{q} _{\mathrm{n}2} \cdot \boldsymbol{r} _{1} +
                           \boldsymbol{q} _{\mathrm{n}1} \cdot \boldsymbol{r} _{2} ) \right ]  \} \, , \\
  \psi _{\mathrm{n} 1} (\boldsymbol{r}_{1}, \boldsymbol{r}_{2})
  &=
  \frac{1}{\sqrt{2}} 
     \{ 
      \exp \left [ i (\boldsymbol{q} _{\mathrm{n}1} \cdot \boldsymbol{r} _{1} +
                           \boldsymbol{q} _{\mathrm{n}2} \cdot \boldsymbol{r} _{2} ) \right ] \notag \\
   &\hspace{12mm}
    - \exp \left [ i (\boldsymbol{q} _{\mathrm{n}2} \cdot \boldsymbol{r} _{1} +
                           \boldsymbol{q} _{\mathrm{n}1} \cdot \boldsymbol{r} _{2} ) \right ]  \} \, .
\end{align}
Using these wave function, we can calculate Hadronic matrix elements of 
Eq.~(\ref{eq:amp-dnn-def}) as
\begin{equation}
  \begin{split}
  \langle \text{dn} _{1} | J ^{0} | ^4 \text{He} \rangle
    &= \sqrt{3} \mathcal{M} _{\text{dnn}} \, , \\
  \langle \text{dn} _{1} | J ^{+} | ^4 \text{He} \rangle
    &= - \sqrt{2} g _{\text{A}} \mathcal{M} _{\text{dnn}} \, , \\
  \langle \text{dn} _{1} | J ^{-} | ^4 \text{He} \rangle
    &=    \sqrt{2} g _{\text{A}} \mathcal{M} _{\text{dnn}} \, ,  \\
  \langle \text{dn} _{1} | J ^{3} | ^4 \text{He} \rangle
    &=    \sqrt{2} g _{\text{A}} \mathcal{M} _{\text{dnn}} \, , \\
  \langle \text{dn} _{0} | J ^{+} | ^4 \text{He} \rangle
    &=    \sqrt{2} g _{\text{A}} \mathcal{M} _{\text{dnn}} ^{\prime} \, , \\
  \langle \text{dn} _{0} | J ^{-} | ^4 \text{He} \rangle
    &= - \sqrt{2} g _{\text{A}} \mathcal{M} _{\text{dnn}} ^{\prime} \, , \\
  \langle \text{dn} _{0} | J ^{3} | ^4 \text{He} \rangle
    &= - \sqrt{2} g _{\text{A}} \mathcal{M} _{\text{dnn}} ^{\prime} \, ,
  \end{split}
\end{equation}
\begin{widetext}
where $\mathcal{M} _{\text{dnn}}$ and $\mathcal{M} _{\text{dnn}} ^{\prime}$ are defined as 
\begin{align}
 \mathcal{M} _{\text{dnn}} 
   &= \left ( \frac{32 \pi ^2 a _{\text{d}}}{a _{\text{He}} (a _{\text{He}} + a _{\text{d}}) ^2} \right ) 
        ^{\frac34} \notag \biggl[
      \exp \left ( - \frac{4 \boldsymbol{q} _{\text{n}2} ^2 
                                           +4 \boldsymbol{q} _{\text{n}2} \cdot \boldsymbol{q} _{\text{d}} 
                                           +3 \boldsymbol{q} _{\text{d}} ^2}
                                           {8 a _{\text{He}}} \right ) 
             -\exp \left ( - \frac{4 \boldsymbol{q} _{\text{n}1} ^2 
                                           +4 \boldsymbol{q} _{\text{n}1} \cdot \boldsymbol{q} _{\text{d}} 
                                           +3 \boldsymbol{q} _{\text{d}} ^2}
                                           {8 a _{\text{He}}} \right ) \biggr]\, , \\
 \mathcal{M} _{\text{dnn}} ^{\prime} 
     &=\frac{1}{\sqrt{2}} 
           \left ( \frac{32 \pi ^2 a _{\text{d}}}
                             {a _{\text{He}} (a _{\text{He}} + a _{\text{d}}) ^2} \right ) 
      ^{\frac{3}{4}} \notag \biggl[ 
       \exp \left ( - \frac{4 \boldsymbol{q} _{\text{n}2} ^2 
                                   +4 \boldsymbol{q} _{\text{n}2} \cdot \boldsymbol{q} _{\text{d}} 
                                   +3 \boldsymbol{q} _{\text{d}} ^2}
                                   {8 a _{\text{He}}} \right ) 
       +\exp \left ( -\frac{4 \boldsymbol{q} _{\text{n}1} ^2 
                                   +4 \boldsymbol{q} _{\text{n}1} \cdot \boldsymbol{q} _{\text{d}} 
                                   +3 \boldsymbol{q} _{\text{d}} ^2}
                                   {8 a _{\text{He}}} \right ) \notag \\
      &\hspace{15mm} 
      -2 \exp \left ( -\frac{3 \boldsymbol{q} _{\text{n}1} ^2 
                                      +2 \boldsymbol{q} _{\text{n}1} \cdot \boldsymbol{q} _{\text{n}2} 
                                      +3 \boldsymbol{q} _{\text{n}2} ^2}
                                   {8 a _{\text{He}}} 
                            -\frac18 \frac{ ( \boldsymbol{q} _{\text{d}}
                                                     +\boldsymbol{q} _{\text{n}1}
                                                     +\boldsymbol{q} _{\text{n}2} ) ^2}
                                   {a _{\text{He}} + a _{\text{d}}} \right ) \biggr] \, .
\end{align}
\end{widetext}

\subsubsection{Amplitude} 

Combining the hadronic matrix elements with the leptonic matrix elements, 
we obtain the square amplitude of Eq.~(\ref{eq:amp-dnn-def}) as 
\begin{align}
   & 
   |\mathcal{M} \bigl( (\tilde \tau ^4\text{He}) \to 
   \tilde \chi_1^0 \nu_\tau \text{dnn} \bigr) |^2 \notag \\
   &= 
   | \langle \text{dnn} 
     |J ^0| ^4\text{He} \rangle | ^2 ~ 
     | \langle \tilde \chi_1^0 \hspace{0.5mm} \nu_\tau 
     |j _0| \tilde \tau \rangle | ^2 \notag \\
   &+ 
   | \langle \text{dnn} 
    |J ^-| ^4\text{He} \rangle | ^2 ~ 
    | \langle \tilde \chi_1^0 \hspace{0.5mm} \nu_\tau 
    |j _+| \tilde \tau \rangle | ^2\notag \\ 
   &+ 
   | \langle \text{dnn} 
     |J ^+| ^4\text{He} \rangle | ^2 ~ 
     | \langle \tilde \chi_1^0 \hspace{0.5mm} \nu_\tau 
     |j _-| \tilde \tau \rangle | ^2 \notag \\
   &+ 
   | \langle \text{dnn} 
     |J ^3| ^4\text{He} \rangle | ^2 ~ 
     | \langle \tilde \chi_1^0 \hspace{0.5mm} \nu_\tau 
     |j _3| \tilde \tau \rangle | ^2 \notag \\
   &= 
   \frac{12 m _{\tilde \chi _1 ^0} G _{\text{F}} ^2 |g _{\text{R}}| ^2}{m _{\tau} ^2} 
      ((1 + 2 g _{\text{A}} ^2 ) \mathcal{M} _{\text{dnn}} ^2 + 2 g _{\text{A}} ^2 
      \mathcal{M} _{\text{dnn}} ^{\prime 2}) E _{\nu} \, .
\end{align}

\subsection{$^4\text{He} \to \text{pnnn}$} 

The matrix element for the $^4\text{He} \to \text{pnnn}$ process is 
\begin{equation}
\begin{split}
   &
   \mathcal{M} (^4\text{He} \to \text{pnnn})
   \\ &=
   \langle \tilde \chi_1^0 \nu_\tau |j_{\mu}| \tilde{\tau} \rangle
   \langle \text{pnnn} |J^{\mu}| ^4\text{He} \rangle \ .
   \label{eq:amp-pnnn-def}
\end{split}  
\end{equation}
The calculation is also performed with an identical step as that of other 
$^4\text{He}$ spallation processes.

\subsubsection{Hadronic matrix element}

The final state of the process is a system composed of a proton and three 
neutrons, and two types of the systems could be brought; (1) $S=0$ and 
$S_z=0$ via vector current (2) $S=1$ and $S_z= \{-1$, $0$, $+1 \}$ 
via axial vector current.

The spin and isospin part of the system for $S=0$ and $S_z=0$ is given by 
\begin{equation}
\begin{split}
   &
   |\text{pnnn} (S=0, S_z=0) \rangle 
   = \frac{1}{4 \sqrt{3}} \epsilon_{ijkl} 
   \\& \times
   \Bigl[ \hspace{0.5mm}
   |\text{n} \uparrow \rangle_i 
   |\text{n} \uparrow \rangle_j 
   |\text{n} \downarrow \rangle_k 
   |\text{p} \downarrow \rangle_l
   + 
   |\text{n} \downarrow \rangle_i 
   |\text{n} \downarrow \rangle_j 
   |\text{n} \uparrow \rangle_k 
   |\text{p} \uparrow \rangle_l 
   \Bigr] , 
\end{split}  
\end{equation}
where we set $\epsilon_{ijkl} = +1$. Here indices $i$, $j$, $k$, $l$ are 
corresponding to spacial coordinates, i.e., spacial part of the wave function 
for $|\text{n} \uparrow \rangle_i  |\text{n} \uparrow \rangle_j 
|\text{n} \downarrow \rangle_k  |\text{p} \downarrow \rangle_l$ is 
\begin{equation}
\begin{split}
   & 
   \psi (\bm{r}_i , \bm{r}_j , \bm{r}_k, \bm{r}_l) 
   \\& \equiv 
   \exp 
   \Bigl[ i ( \bm{q}_\text{n1} \cdot \bm{r}_i
   + \bm{q}_\text{n2} \cdot \bm{r}_j
   + \bm{q}_\text{n3} \cdot \bm{r}_k
   + \bm{q}_\text{p} \cdot \bm{r}_l ) \Bigr] .
\end{split}  
\end{equation}
Here each spacial part of proton and neutron are simply taken to be 
plane wave form. 
Similarly, the spin and isospin parts of the system for $S=1$ and $S_z= 
\{-1$, $0$, $+1 \}$ are given by
\begin{equation}
\begin{split} 
   &
   |\text{pnnn} (S=1, S_z=+1) \rangle 
   \\&= 
   \frac{1}{2 \sqrt{6}} \epsilon_{ijkl}
   |\text{n} \uparrow \rangle_i
   |\text{n} \uparrow \rangle_j
   |\text{n} \downarrow \rangle_k
   |\text{p} \uparrow \rangle_l
\end{split}  
\end{equation}
\begin{equation}
\begin{split} 
   &
   |\text{pnnn} (S=1 ,S_z=0) \rangle 
   = 
   \frac{1}{4 \sqrt{3}} \epsilon _{ijkl} 
   \\& \times
   \Bigl[  
   |\text{n} \uparrow \rangle_i 
   |\text{n} \uparrow \rangle_j
   |\text{n} \downarrow \rangle_k 
   |\text{p} \downarrow \rangle_l 
   - 
   |\text{n} \downarrow \rangle_i 
   |\text{n} \downarrow \rangle_j
   |\text{n} \uparrow \rangle_k 
   |\text{p} \uparrow \rangle_l  \Bigr]
\end{split}  
\end{equation}
\begin{equation}
\begin{split} 
   &
   |\text{pnnn} (S=1, S_z=-1) \rangle 
   \\&= 
   \frac{1}{2\sqrt{6}} \epsilon_{ijkl}
   |\text{n} \downarrow \rangle_i
   |\text{n} \downarrow \rangle_j
   |\text{n} \uparrow \rangle_k
   |\text{p} \downarrow \rangle_l
\end{split}  
\end{equation}
Spatial parts of them are same as the system for $S=0$ and $S_z=0$.

The hadronic matrix element in (\ref{eq:amp-pnnn-def}) is calculated 
with built wave functions and explicit form of each current as follows, 
\begin{equation}
\begin{split}
   &
   \langle \text{pnnn} |J^0| ^4\text{He} \rangle 
   = \sqrt{2} \mathcal{M}_\text{pnnn} , 
   \\&
   \langle \text{pnnn} |J^+| ^4\text{He} \rangle 
   = - \sqrt{2} g_\text{A} \mathcal{M}_\text{pnnn}  , 
   \\&
   \langle \text{pnnn} |J^+| ^4\text{He} \rangle 
   = \sqrt{2} g_\text{A} \mathcal{M}_\text{pnnn}  , 
   \\&
   \langle \text{pnnn} |J^3| ^4\text{He} \rangle 
   = \sqrt{2} g_\text{A} \mathcal{M}_\text{pnnn} , 
\end{split}
\end{equation}
\begin{widetext}
where $\mathcal{M} _{\text{pnnn}}$ is define as follow: 
\begin{equation}
\begin{split}
   \mathcal{M}_\text{pnnn} 
   &= 
   \biggl( \frac{32 \pi^3}{a_\text{He}^3} \biggr)^{3/4} 
   \biggl\{ \exp \biggl[ 
   - \frac{1}{2 a_\text{He}} \bigl( 
   \bm{q}_\text{n2}^2 + 
   \bm{q}_\text{n3}^2 + 
   \bm{q}_\text{p}^2 +  
   \bm{q}_\text{n2} \cdot \bm{q}_\text{n3} + 
   \bm{q}_\text{n2} \cdot \bm{q}_\text{p} + 
   \bm{q}_\text{n3} \cdot \bm{q}_\text{p}  \bigr)
   \biggr] 
   \\& ~~ - \exp \biggl[ 
   - \frac{1}{2 a_\text{He}} \bigl( 
   \bm{q}_\text{n1}^2 + 
   \bm{q}_\text{n3}^2 + 
   \bm{q}_\text{p}^2 + 
   \bm{q}_\text{n1} \cdot \bm{q}_\text{n3} + 
   \bm{q}_\text{n1} \cdot \bm{q}_\text{p} + 
   \bm{q}_\text{n3} \cdot \bm{q}_\text{p}  \bigr)
   \biggr]
   \biggr\}
\end{split}
\end{equation}
\end{widetext}

\subsubsection{Amplitude} 

Combining the hadronic matrix elements with the leptonic matrix elements, 
we obtain the square amplitude of Eq.~(\ref{eq:amp-pnnn-def}) as 
\begin{equation}
\begin{split}
   & 
   |\mathcal{M} \bigl( (\tilde \tau ^4\text{He}) \to 
   \tilde \chi_1^0 \nu_\tau \text{pnnn} \bigr) |^2 
   \\&= 
   |\langle \text{pnnn} |J^0| ^4\text{He} \rangle|^2 \  
   |\langle \tilde \chi_1^0  \nu_\tau |j_0| \tilde \tau \rangle|^2  
   \\&+ 
   |\langle \text{pnnn} |J^-| ^4\text{He} \rangle|^2 \  
   |\langle \tilde \chi_1^0 \nu_\tau |j_+| \tilde \tau \rangle|^2  
   \\&+ 
   |\langle \text{pnnn} |J^+| ^4\text{He} \rangle|^2 \ 
   |\langle \tilde \chi_1^0 \nu_\tau |j_-| \tilde \tau \rangle|^2  
   \\&+ 
   |\langle \text{pnnn} |J^3| ^4\text{He} \rangle|^2 \ 
   |\langle \tilde \chi_1^0 \nu_\tau |j_3| \tilde \tau \rangle|^2  
   \\&= 
   \frac{8 m_{\tilde \chi_1^0} G_\text{F}^2 |g_\text{R}|^2}{m_\tau^2} 
   (1 + 3g_\text{A}^2 ) \mathcal{M}_\text{pnnn}^2  E_\nu  .
\end{split}
\end{equation}


\end{document}